\documentclass[reprint,aps,showpacs,pra,nofootinbib]{revtex4-1}
  \usepackage{newlfont}
  \usepackage{graphicx}
  \usepackage{amssymb}
  \usepackage{amsmath}
  \usepackage{bbm}
  \usepackage{braket}
  \usepackage[varg]{txfonts}

\newcommand{\be}{\begin{equation}}
\newcommand{\ee}{\end{equation}}
\newcommand{\beq}{\begin{eqnarray}}
\newcommand{\eeq}{\end{eqnarray}}
\newcommand{\Tr}{\text{Tr}}

\begin{document}

\title{Tripartite realism-based quantum nonlocality}

\author{D. M. Fucci}
\author{R. M. Angelo}
\affiliation{Department of Physics, Federal University of Paran\'a, P.O. Box 19044,
81531-980 Curitiba, Paran\'a, Brazil}

\begin{abstract}
From an operational criterion of physical reality, a quantifier of realism-based nonlocality was recently introduced for two-part quantum states. This measure has shown to capture aspects that are rather different from Bell nonlocality. Here we take a step further and introduce a tripartite realism-based nonlocality quantifier. We show that this measure reduces to genuine tripartite entanglement for a certain class of pure tripartite states and manifests itself in correlated mixed states even in the absence of quantum correlations. A case study for noisy GHZ and W states points out the existence of scenarios where the realism-based nonlocality is monogamous.
\end{abstract}

\maketitle

\section{Introduction} 
One of the most astonishing aspects of nature is unveiled by Bell's theorem, which proves that the predictions of quantum mechanics are incompatible with physical theories relying upon the local causality hypotesis~\cite{Bell1964}, which is mathematically formulated as $p(a,b|A,B)=\sum_{\lambda} p_{\lambda}\,p(a|A,\lambda)\,p(b|B,\lambda)$. This relation assumes that the joint probability $p(a,b|A,B)$ of finding the outcomes $a$ and $b$ in measurements of observables $A$ and $B$, respectively, is factorable into local probability distributions, $p(a|A,\lambda)$ and $p(b|B,\lambda)$, by virtue of the specification of a hidden variable $\lambda$ governed by a probability distribution $p_{\lambda}$. Today, it is a fact firmly backed by solid experimental evidence~\cite{Hensen2015, Giustina2015, Shalm2015, Hensen2016, Rosenfeld2017, Handsteiner2017, Rauch2018} that nature does not submit itself to such hypothesis. This suggests that ``spooky actions'' may take place at distance, a phenomenon known as Bell nonlocality~\cite{Brunner2014}.

Quantum mechanics provides accurate accounts for experimental data. On the one hand, it is known that separable states $\rho_s=\sum_{\lambda}p_{\lambda}\,\rho_{\lambda}^{\mathcal{A}}\otimes\rho_{\lambda}^{\mathcal{B}}$ on $\mathcal{H_A\otimes H_B}$ produce distributions $p(a,b|A,B)=\Tr(A_{a}\otimes B_{b}\,\rho_{s})$ admitting the factorization prescribed by the local causality hypothesis, for projectors $A_a$ on $\mathcal{H_A}$ and $B_b$ on $\mathcal{H_B}$. In this case, no Bell inequality is violated and the $\rho_s$ is said Bell local. On the other hand, quantum mechanics predicts that all entangled pure states are Bell nonlocal~\cite{Gisin1991,Yu2012}. For mixed states, Bell nonlocality is known to demand entanglement, while the converse is not true~\cite{Brunner2014}, which implies that the class of Bell nonlocal states form a subset of the entangled states. The existence of maximally entangled states that are not maximally Bell nonlocal~\cite{Acin2002, Methot2007, Vidick2011, Camalet2017}---the so called anomaly---reveals how tricky the quantification of Bell nonlocality may be. In effect, unlike entanglement quantification~\cite{Horodecki2009}, the task of quantifying Bell nonlocality still elicits debate. Some of the recent approaches make reference to maximal violations of Bell inequalities~\cite{CA2016}, performances in communication tasks~\cite{Maudlin1992, Brassard1999, Steiner2000, Bacon2003, Branciard2011}, noise resistance~\cite{Kaszlikowski2000, Laskowski2014}, and the volume of violation~\cite{Fonseca2015}, the latter solving the anomaly problem for specific states.

Even though Bell's original argument makes no formal link with realism~\cite{Gisin2012}, one can argue that this concept---actually the absence of it---is mandatory for the manifestation of nonlocality. The example given in Ref.~\cite{BA2015} illustrates this point. Suppose that two qubits share a singlet state, that is, a physical preparation that constraints the total spin as $\mathfrak{s}_z=\mathfrak{s}_z^\mathcal{A}+\mathfrak{s}_z^\mathcal{B}=0$. The definiteness of the total spin allows us to say that it is an element of reality, even though the spins of the parts are not. This conservation law is preserved as the spins are sent to far distant locations. Now, if $\mathfrak{s}_z^\mathcal{A}$ is measured and becomes an element of reality, the conservation law forces $\mathfrak{s}_z^\mathcal{B}$ to instantaneously become so. Since the emergence of reality in part $\mathcal{B}$ is then causally induced by a remote disturbance, some aspect of nonlocality is presumed to take place. Inspired by the lack of such a clear link between realism and violations of the local causality hypothesis, a notion of nonlocality has recently been put forward that makes explicit use of a criterion of realism~\cite{BA2015}. This novel aspect of nonlocality counts with a nonanomalous quantifier, is remarkably more resilient to noise than other quantum resources, and reduces to bipartite entanglement for pure states~\cite{GA2018,GA2019}.

While on the Bell nonlocality side, a significant literature exists concerning multipartite settings~\cite{Svetlichny1987,Bancal2009,Bancal2011,Bancal2013,Rosier2017,Chaves2017}, it is still not clear whether it is possible to extend the notion of realism-based nonlocality even to tripartite scenarios. This work aims at starting this research program. Our basic strategy consists of performing a slight adaptation of the current measure of realism-based nonlocality so as to make it applicable to all possible bipartitions of a tripartite state. Then, we construct a quantifier of genuine tripartite realism-based nonlocality, derive its properties, and prove that it reduces to genuine tripartite entanglement for a given class of pure states. In addition, we discuss whether the introduced measure is monogamous.

\section{Preliminary concepts} 
Here we present a brief review of the realism-based nonlocality. As seminally introduced by Einstein, Podolsky, and Rosen (EPR)~\cite{EPR1935}, the concept of ``elements of reality'' was related with the idea of full predictability without disturbance. This criterion, however, fails to diagnose situations where the elements of reality cannot be predicted because of mere subjective ignorance. For instance, if Alice measures the $z$-spin of a particle but does not let Bob know the outcome, one would not say that the measured spin is not an element of reality just because Bob cannot predict the result before the next measurement on this particle. To overcome this difficulty, Bilobran and Angelo introduced an operational criterion to identify elements of reality for a generic preparation $\rho$ on $\mathcal{H_A\otimes H_B}$~\cite{BA2015}. Their key premise is that once a discrete-spectrum observable $A=\sum_aaA_a$ is measured, there must be an element of reality associated with $A$, even when the measurement outcome is not revealed. The post-measurement state reads
\be
\sum_a(A_a\otimes\mathbbm{1}_\mathcal{B})\,\rho\,(A_{a}\otimes\mathbbm{1}_\mathcal{B}) = \sum_{a}p_{a}A_{a} \otimes \rho_{\mathcal{B}|a}=:\Phi_A(\rho),
\ee
where $p_a=\Tr(A_a\otimes\mathbbm{1}_\mathcal{B}\,\rho)$, $\rho_{\mathcal{B}|a}=\Tr_\mathcal{A}(A_a\otimes\mathbbm{1}_\mathcal{B}\,\rho)/p_a$, and $A_aA_{a'}=A_a\delta_{aa'}$. The unrevealed measurement of $A$ is henceforth denoted by $\Phi_A$, which is a completely positive trace-preserving unital map. Bilobran and Angelo then take $\Phi_A(\rho)=\rho$ as a criterion of realism with basis on the following rationale: if measuring $A$ and not revealing the outcome do not effectively change the state of the system, then this state is just an epistemic description of the part $\mathcal{A}$, where $A$ is already an element of reality. As expected, one has $\Phi_A\Phi_A(\rho)=\Phi_A(\rho)$, meaning that successive applications of an unrevealed measurement of $A$ over an $A$-reality state does not create ``irreality'' for $A$. It is then proposed the measure 
\be
\mathfrak{I}_A(\rho):=S(\Phi_A(\rho))-S(\rho)
\label{frakI}
\ee
of the irreality of $A$ for a given preparation $\rho$, where $S$ denotes the von Neumann entropy. It can be shown that irreality is nonnegative, vanishes if and only if $\rho=\Phi_A(\rho)$, and is nonincreasing under completely positive trace-preserving maps. By use of $\mathfrak{I}_A$, an information-reality complementarity relation was recently derived for generic unrevealed measurements~\cite{DA2018}, with some results having being experimentally verified through a photonics platform~\cite{Mancino2018}.

In the same work, Bilobran and Angelo propose to use measure \eqref{frakI} to quantify by how much the irreality of an observable $A$ in the site $\mathcal{A}$ changes due to unrevealed measurements of an observable $B$ acting on part $\mathcal{B}$. The authors then introduced the contextual realism-based nonlocality
\be
\eta_{A|B}(\rho):=\mathfrak{I}_A(\rho)-\mathfrak{I}_A(\Phi_{B}(\rho)),
\label{etaAB}
\ee
where $B=\sum_bbB_b$ and $\Phi_B(\rho)=\sum_b(\mathbbm{1}_\mathcal{A}\otimes B_b)\,\rho\,(\mathbbm{1}_\mathcal{A}\otimes B_b)$, for projectors $B_b$. Here the context is defined by the pair $\{A,B\}$. It has been shown that $\eta_{A|B}(\rho) \geq 0$, with equality holding for product states, $\rho=\rho_\mathcal{A}\otimes\rho_\mathcal{B}$, and for states of reality, that is, $\rho =\Phi_A(\rho)$, $\rho=\Phi_B(\rho)$, or $\rho = \Phi_{A,B}(\rho)\equiv\Phi_A\Phi_B(\rho)$. 

By maximizing the contextual realism-based nonlocality over all possible contexts, Gomes and Angelo then introduced the bipartite realism-based nonlocality~\cite{GA2018}
\be
\mathcal{N}_2(\rho) := \underset{\left\{A,B\right\}}{\max}\: \eta_{A|B}(\rho),
\ee
which diagnoses the nonlocality of the state $\rho$ on $\mathcal{H_A\otimes H_B}$. This quantity has shown to be nonanomalous, since for maximally entangled bipartite states, $\ket{\psi}=\sum_{i=1}^{d} \ket{i} \ket{i}/\text{\small $\sqrt{d}$}$, it reduces to the entanglement $E$ of $\ket{\psi}$, that is, $\mathcal{N}_{2}(\varrho)=S(\varrho_\mathcal{R})\equiv E(\varrho)$, where $\mathcal{R\in\{A,B\}}$, $\varrho = \ket{\psi} \bra{\psi}$, and $\varrho_\mathcal{A(B)}=\Tr_\mathcal{B(A)}\varrho$. A distinctive feature of $\mathcal{N}_2$ can be readily appreciated for the classical-classical state $\rho_\text{cc} = \sum_{i} p_{i} A'_{i} \otimes B'_{i}$, where $A'=\sum_i a'_iA'_i$ and $B'=\sum_ib'_iB'_i$. As shown in Ref.~\cite{GA2019} this state has none of the well-established nonclassical features, namely, Bell nonlocality, EPR steering, entanglement, and quantum discord. Still, by choosing a context $\{A,B\}$ maximally incompatible with $\{A',B'\}$ we find $\mathcal{N}_2(\rho_\text{cc})=\eta_{A|B}(\rho_\text{cc})=H(\{p_i\})>0$, where $H(\left\{p_{i}\right\})$ is the Shannon entropy of the probability distribution $p_i$. Since $\eta_{A'|B'}(\rho_\text{cc})=0$, one sees that $\mathcal{N}_2$ is able to capture aspects of incompatibility which suffice to produce realism changes at distance. 

\section{Tripartite realism-based nonlocality} 
With basis on Bilobran and Angelo's approach to realism and nonlocality, we now construct the notion of genuine tripartite realism-based nonlocality. Hereafter, for the sake of notational simplicity, we reserve the term nonlocality for referring to the bi- and tripartite versions of the reaslim-based nonlocality, in distinction to Bell nonlocality. Consider a tripartite preparation $\rho \in \mathcal{H}_{\mathcal{A}} \otimes \mathcal{H}_{\mathcal{B}} \otimes \mathcal{H}_{\mathcal{C}}$ such that the parts of the system are sent to distinct laboratories, $\mathcal{A}$, $\mathcal{B}$, and $\mathcal{C}$, which are far apart from each other. Let $R\in\{A,B,C\}$ on $\mathcal{H_R}$ be an observable accessible only in the laboratory $\mathcal{R\in\{A,B,C\}}$. We then introduce the bipartite contextual nonlocality
\be
\eta_{A|B,C}(\rho) := \mathfrak{I}_A(\rho) - \mathfrak{I}_A(\Phi_{B,C}(\rho)),
\label{eta_A|BC}
\ee
where $\Phi_{B,C}=\Phi_B\Phi_C$. This formula applies for any permutations of $A$, $B$, and $C$. Being a natural extension of definition \eqref{etaAB}, $\eta_{A|B,C}$ captures changes in the reality of $A$ given that local unrevealed measurements $\Phi_B$ and $\Phi_C$ are conducted in the remote labs $\mathcal{B}$ and $\mathcal{C}$. Because irreality never increases upon completely positive trace-preserving maps, it follows that $\eta_{A|B,C}(\rho)\geq 0$, the equality applying for states of reality $\rho=\Phi_{A}(\rho)$ and $\rho=\Phi_{B,C}(\rho)$, for the full realism state $\text{\small $\bigotimes$}_\mathcal{R=A,B,C}\tfrac{\mathbbm{1}_\mathcal{R}}{d_\mathcal{R}}\equiv\tfrac{\mathbbm{1}}{d}$, where $d=d_\mathcal{A}d_\mathcal{B}d_\mathcal{C}$, and for uncorrelated states as $\rho=\rho_\mathcal{A}\otimes\rho_\mathcal{BC}$. From this we see that irreality and correlations are prerequisites for nonlocality. It is also noteworthy that $\eta_{A|B,C}(\rho_{\mathcal{AB}}\otimes\rho_\mathcal{C})=\eta_{A|B}(\rho_\mathcal{AB})$, which is desirable since part $\mathcal{C}$ is fully irrelevant in this case. The same applies if $\mathcal{B}$ is uncorrelated.

By maximizing over all trios $\{A,B,C\}$ of local observables, we introduce the amount of nonlocality $\mathcal{N}_\mathcal{A|BC}$ associated with realism changes in part $\mathcal{A}$ induced by local measurements in parts $\mathcal{B}$ and $\mathcal{C}$ for a given preparation $\rho$, that is,
\be
\mathcal{N}_{\mathcal{A}|\mathcal{BC}}(\rho) := \underset{\left\{A,B,C\right\}}{\max}\: \eta_{A|B,C}(\rho).
\label{N_A|BC}
\ee
When $\mathcal{N}_{\mathcal{A}|\mathcal{BC}}(\rho) > 0$ we are certain that there exists at least one setting $\{A,B,C\}$ through which one is able to spot a change in the reality of $A$ when $B$ and $C$ are measured. It can be verified that $\mathcal{N}_{\mathcal{A}|\mathcal{BC}}(\rho)$ vanishes when $\rho=\mathbbm{1}/d$ or $\rho=\rho_\mathcal{A}\otimes\rho_\mathcal{BC}$. In addition, one has $\mathcal{N}_\mathcal{A|BC}(\rho_\mathcal{AB}\otimes\rho_\mathcal{C})=\mathcal{N}_2(\rho_\mathcal{AB})$.

Now, to build our quantifier of genuine tripartite nonlocality, we invoke Bennett {\it et al.}'s proposal~\cite{Bennett2011} for the identification of genuine multipartite correlations: ``A state of $n$ particles has genuine $n$-partite correlations if it is nonproduct in every bipartite cut''. As shown by Ma \textit{et al.}~\cite{Ma2011}, this approach is able to produce, for instance, a measure of genuine tripartite entanglement for pure states, namely,
\be
E_{3}(\rho) := \min \Big\{ E_{\mathcal{A}|\mathcal{BC}}(\rho), E_{\mathcal{B}|\mathcal{AC}}(\rho), E_{\mathcal{C}|\mathcal{AB}}(\rho) \Big\}, 
\label{E3}
\ee
where $E_\mathcal{A|BC}(\rho)=S(\rho_\mathcal{A})$ is the entanglement entropy of the partition $\mathcal{A|BC}$, with a similar interpretation for the other terms. Being different from quantum discord, $\mathcal{N_{A|BC}}$ cannot be termed a strict measure of quantum correlations. Still, since it makes reference to nonlocal connections between the parts, the above rationale involving bipartite cuts is applicable. This allows us to introduce the quantifier of genuine tripartite realism-based nonlocality:
\be
\mathcal{N}_{3}(\rho) := \min \Big\{ \mathcal{N}_{\mathcal{A}|\mathcal{BC}}(\rho),\mathcal{N}_{\mathcal{B}|\mathcal{AC}}(\rho),\mathcal{N}_{\mathcal{C}|\mathcal{AB}}(\rho) \Big\}.
\label{N3}
\ee
From what we have for measure \eqref{N_A|BC}, it follows that $\mathcal{N}_3$ vanishes for states like $\mathbbm{1}/d$ and $\rho_\mathcal{A}\otimes\rho_\mathcal{BC}$ (including states with permuted indexes).

We now prove an interesting result with respect to the special class of pure states $\ket{\varphi}$ admitting a Schmidt decomposition $\ket{\varphi} = \sum_i\sqrt{\xi_i} \ket{\alpha_i}\ket{\beta_i}\ket{\gamma_i}$~\cite{Pati2000}, for bases connected with the observables $\alpha=\sum_i\alpha_i\ket{\alpha_i}\bra{\alpha_i}$, $\beta=\sum_i \beta_i\ket{\beta_i}\bra{\beta_i}$, and $\gamma=\sum_i\gamma_i\ket{\gamma_i}\bra{\gamma_i}$ acting on respective spaces $\mathcal{H_{A,B,C}}$. We start by plugging definition \eqref{frakI} into the contextual nonlocality, \eqref{eta_A|BC}, with $\varsigma=\ket{\varphi}\bra{\varphi}$, to write
\be\label{eq1}
\eta_{A|B,C}(\varsigma) = S(\Phi_{A}(\varsigma)) + S(\Phi_{B,C}(\varsigma)) - S(\Phi_{A,B,C}(\varsigma)) - S(\varsigma).
\ee
For pure states, $S(\varsigma) = 0$. Given the entropy monotonicity, $S(\Phi_{R}(\varsigma)) \geq S(\varsigma)$, which implies that $S(\Phi_{R,Q}(\varsigma)) \geq S(\Phi_{R}(\varsigma))$ and $S(\Phi_{A,B,C}(\varsigma)) \geq S(\Phi_{R,Q}(\varsigma))$, with $R,Q\in\{A,B,C\}$, we find
\be\label{eq2}
2S(\Phi_{A,B,C}(\varsigma)) \geq S(\Phi_{A}(\varsigma)) + S(\Phi_{B,C}(\varsigma)),
\ee
which saturates when $\Phi_{A,B,C}(\varsigma)=\Phi_{B,C}(\varsigma)=\Phi_A(\varsigma)$. Combining this inequality with Eq.~\eqref{eq1} gives
\be\label{eq3}
\eta_{A|B,C}(\varsigma) \leq \tfrac{1}{2} \Big[S(\Phi_{A}(\varsigma)) + S(\Phi_{B,C}(\varsigma))\Big].
\ee
Therefore, the maximization of $\eta_{A|B,C}(\varsigma)$ will come by the saturation of this inequality. This is accomplished by the choice $A=\alpha$, $B=\beta$, and $C=\gamma$, which can be shown to give $\Phi_{\alpha}(\varsigma) = \Phi_{\beta,\gamma}(\varsigma) = \Phi_{\alpha,\beta,\gamma}(\varsigma) = \sum_{i} \xi_{i} \ket{\alpha_{i}}\bra{\alpha_{i}} \otimes \ket{\beta_{i}}\bra{\beta_{i}} \otimes \ket{\gamma_{i}}\bra{\gamma_{i}}$. Hence, $\mathcal{N}_\mathcal{A|BC}(\varsigma)=\eta_{\alpha|\beta,\gamma}(\varsigma)=S(\Phi_{\alpha}(\varsigma))=H(\{\xi_i\})$. Given the symmetry of $\ket{\varphi}$, one does not expect different results for the other bipartitions, so that $\mathcal{N}_{3}(\varsigma)=H(\{\xi_i\})$. The entanglement of the cut $\mathcal{A|BC}$ is given by $E_{\mathcal{A}|\mathcal{BC}}(\varsigma)=S(\Tr_\mathcal{A}\varsigma)=H(\{\xi_i\})$. Using symmetry again and definition~\eqref{E3}, we obtain $E_3(\varsigma)=H(\{\xi_i\})$, which finally gives
\be 
\mathcal{N}_3(\varsigma)=E_3(\varsigma) \qquad\quad \big(\varsigma=\ket{\varphi}\bra{\varphi}\big).
\ee 
This implies that, as far as the class $\ket{\varphi}$ of tripartite pure states is concerned, $\mathcal{N}_3$ is surely nonanomalous with respect to genuine tripartite entanglement as measured by the formula~\eqref{E3}. 

Consider now a classical-classical-classical state, defined as $\rho_\text{ccc} = \sum_ip_iA'_i\otimes B'_i\otimes C'_i$, with $A' = \sum_ia'_iA'_i$, $B'=\sum_ib'_iB'_i$, and $C'=\sum_ic'_iC'_i$. It is clear that this is a separable state, thus possessing no form of entanglement. Yet, it does present tripartite nonlocality. To see this, let the context $\{A,B,C\}$ be maximally incompatible with $\{A',B',C'\}$. Since $\Phi_A(A_i')=\mathbbm{1}_\mathcal{A}/d_\mathcal{A}$, with similar relations for the other parts, direct calculations give $\eta_{A|B,C}(\rho_\text{ccc})=H(\{p_i\})$. This guarantees that $\mathcal{N_{A|BC}}(\rho_\text{ccc})>0$ and, via symmetry, that $\mathcal{N}_3(\rho_\text{ccc})>0$. In other words, just as $\mathcal{N}_2$, tripartite nonlocality may manifest itself even when no quantum correlation is present.

\section{Applications} 
\begin{figure}[htb]
\centering
\includegraphics[scale=0.34]{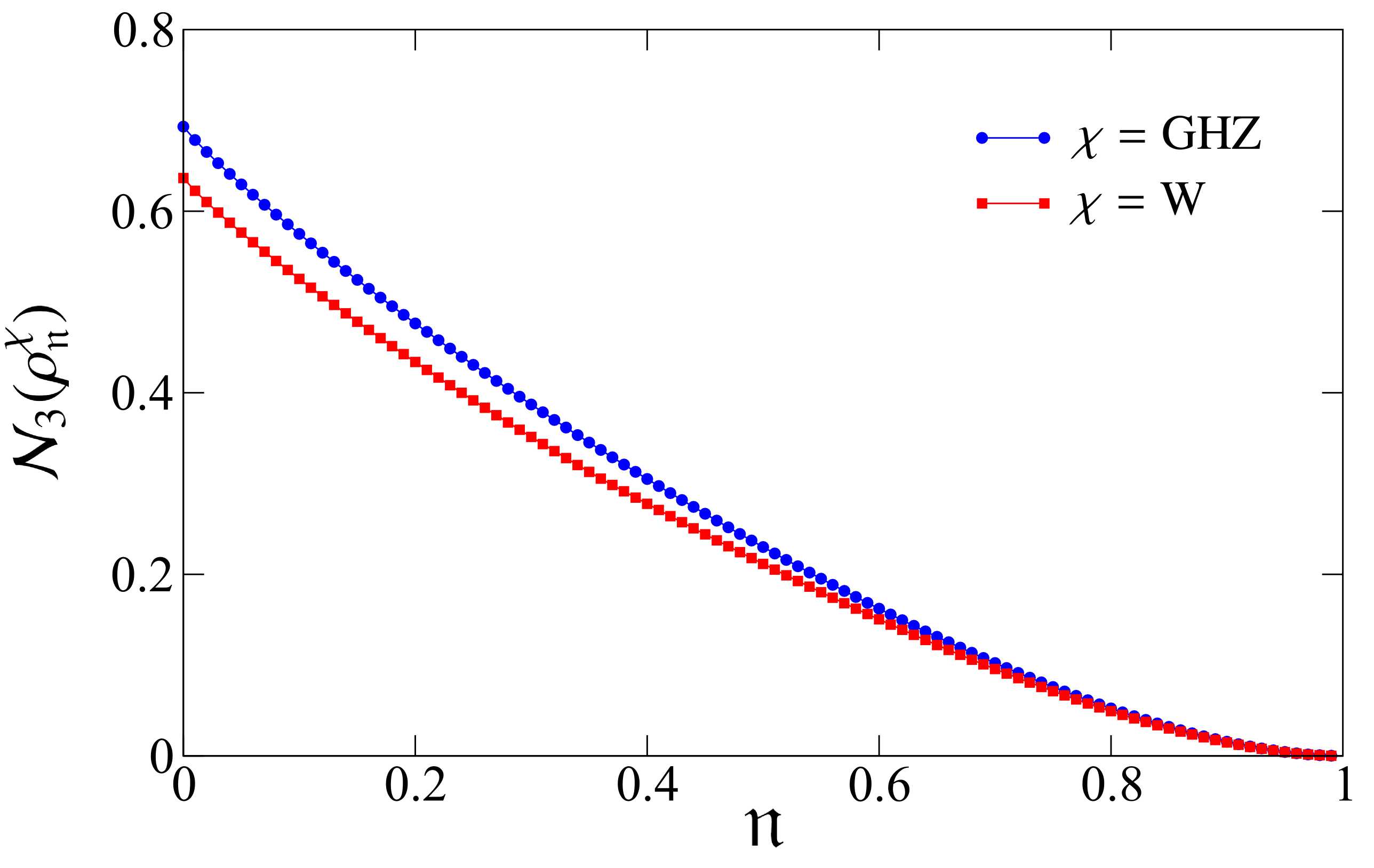}
\caption{\small Genuine tripartite nonlocality $\mathcal{N}_{3}(\rho_\mathfrak{n}^{\chi})$ for the noisy three-qubit states \eqref{rho_n} as a function of the noise amount $\mathfrak{n}$. Blue circles correspond to the noisy GHZ state ($\chi=\text{GHZ}$) and red squares to the noisy W state ($\chi=\text{W}$). Tripartite nonlocality monotonically decreases with noise.}
\label{fig1}
\end{figure}

Here we calculate $\mathcal{N}_3$ for the noisy three-qubit states
\be
\rho_\mathfrak{n}^{\chi}:=\mathfrak{n}\,\tfrac{\mathbbm{1}}{8}+(1-\mathfrak{n})\ket{\Psi_{\chi}}\bra{\Psi_{\chi}},
\label{rho_n}
\ee 
where $\mathfrak{n}\in[0,1]$ gives the noise (or impurity) added to the tripartite pure state $\ket{\Psi_{\chi}}$, which, with $\chi\in\{\text{GHZ,W}\}$, assumes
\begin{subequations}
\beq 
\ket{\Psi_\text{\tiny GHZ}}&=&\tfrac{1}{\sqrt{2}}\Big(\ket{000}+\ket{111}\Big), \\
\ket{\Psi_{\text{\tiny W}}}&=&\tfrac{1}{\sqrt{3}}\Big(\ket{100}+\ket{010}+\ket{001} \Big).
\eeq 
\end{subequations}
Even for these relatively simple states, the evaluation of the genuine tripartite nonlocality $\mathcal{N}_3$ is a hard computational problem due to the optimizations demanded by definitions \eqref{N_A|BC} and \eqref{N3}. Incidentally, due to the subsystem-permutation symmetry, we have $\mathcal{N}_3=\mathcal{N_{A|BC}=N_{B|AC}=N_{C|AB}}$, which simplifies our task and makes it numerically feasible. To obtain $\mathcal{N}_\mathcal{A|BC}$ via maximization of $\eta_{A|B,C}$, we considered spin operators $A=\hat{a}\cdot\vec{\sigma}$ with $\hat{a}=(\sin\theta_a\cos\varphi_a,\sin\theta_a\sin\varphi_a,\cos\theta_a)$ and $\vec{\sigma}=(\sigma_x,\sigma_y,\sigma_z)$, for Pauli matrices $\sigma_{x,y,z}$. Similarly, we parametrized $B$ and $C$ with unit vectors $\hat{b}$ and $\hat{c}$, respectively, thus getting the angles $\theta_{a,b,c}\in[0,\pi]$ and $\varphi_{a,b,c}\in[0,2\pi]$, over which the optimization were performed. We then defined a grid by letting these angles vary in their domain with increments of $\pi/8$, which yielded a set of 2,985,984 distinct settings $\{A,B,C\}$. We computed $\eta_{A|BC}$ for each setting of the set and then picked the maximum. This process was repeated for each value $\mathfrak{n}$ of noise, which ranged in the domain [0,1] with increments of 0.01, and for $\chi=\{\text{GHZ,W}\}$. Our results are presented in Fig.~\ref{fig1}. Such numerical analysis was realized also for $10^6$ randomly generated settings $\{A,B,C\}$ and no appreciable difference was found, meaning that these results are statistically reliable. Throughout our numerical investigations, we found some noteworthy settings. For the GHZ state ($\mathfrak{n}=0$), we found that $\eta_{A|B,C}$ reaches the maximum value $\ln{2}$ when $A=\sigma_z$ and at least one of the observables $B$ and $C$ is equal to $A$. The contextual nonlocality is also maximal when $(A,B,C)$ assumes $(\sigma_x,\sigma_x,\sigma_x)$, $(\sigma_y,\sigma_y,\sigma_x)$, $(\sigma_y,\sigma_x,\sigma_y)$, or $(\sigma_x,\sigma_y,\sigma_y)$, a set of measurement operators that when acting over a GHZ state is known to provide predictions that are in conflict with the local realism hypothesis~\cite{BEZ2000}. For the W state, the maximum contextual nonlocality, $0.6364$, was found for $A=B=C=\sigma_z$.

Notably, the genuine tripartite nonlocality $\mathcal{N}_3$ is a monotonically decreasing function of the noise $\mathfrak{n}$, strictly vanishing only in the scenario of no purity whatsoever ($\mathfrak{n}=1$). This points out the stronger resilience of $\mathcal{N}_3$ in comparison with other measures of nonclassicality which abruptly vanish under high levels of noise. For instance, as far as $\rho_\mathfrak{n}^\text{\tiny GHZ}$ is concerned, it is known that tripartite entanglement disappears for $\mathfrak{n}\geq 4/5$~\cite{Schack2000,Dur2000}, Bell nonlocality for $\mathfrak{n}>1/2$ (with two up to five measurements per site)~\cite{Gruca2010}, and steering for $\mathfrak{n}\gtrsim 0.225$~\cite{Costa2018}. For $\rho_\mathfrak{n}^\text{\tiny W}$, entanglement vanishes for $\mathfrak{n} \geq 0.8220$~\cite{Chen2012}, Bell nonlocality for $\mathfrak{n} > 0.3558$ ($0.3952$) with two (three) measurements per site~\cite{Gruca2010}, and steering for $\mathfrak{n}\gtrsim 0.1634$~\cite{Costa2018}. An equivalent noise resilience was verified for $\mathcal{N}_2$ in Ref.~\cite{GA2018}.

\section{Monogamy of nonlocality} 
A monogamy inequality $Q_\mathcal{A|BC}\geq Q_\mathcal{A|B}+Q_\mathcal{A|C}$ for a given resource measure $Q$ gives an upper bound to the shareability of the related resource among the parts of the system. We now assess whether the genuine tripartite nonlocality $\mathcal{N}_3$ exhibits this property. In fact, we want to test the relation
\be 
\mathcal{N}_3^{\alpha}(\rho_\mathcal{ABC})\geq \mathcal{N}_2^{\alpha}(\rho_\mathcal{AB})+\mathcal{N}_2^{\alpha}(\rho_\mathcal{AC}),
\label{monogamy}
\ee 
where $\rho_\mathcal{AB}$ and $\rho_\mathcal{AC}$ are reduced states, and $\alpha\in\mathbbm{R}_{>0}$ is a parameter intended to give to a further measure $\mathcal{N}_{3}^{\alpha}$, monotonically related with $\mathcal{N}_3$, an extra chance to satisfy monogamy. This strategy has proven successful in establishing monogamy for general measures of nonclassicality~\cite{Jin2019}.

\begin{figure}[b]
\includegraphics[width=\columnwidth]{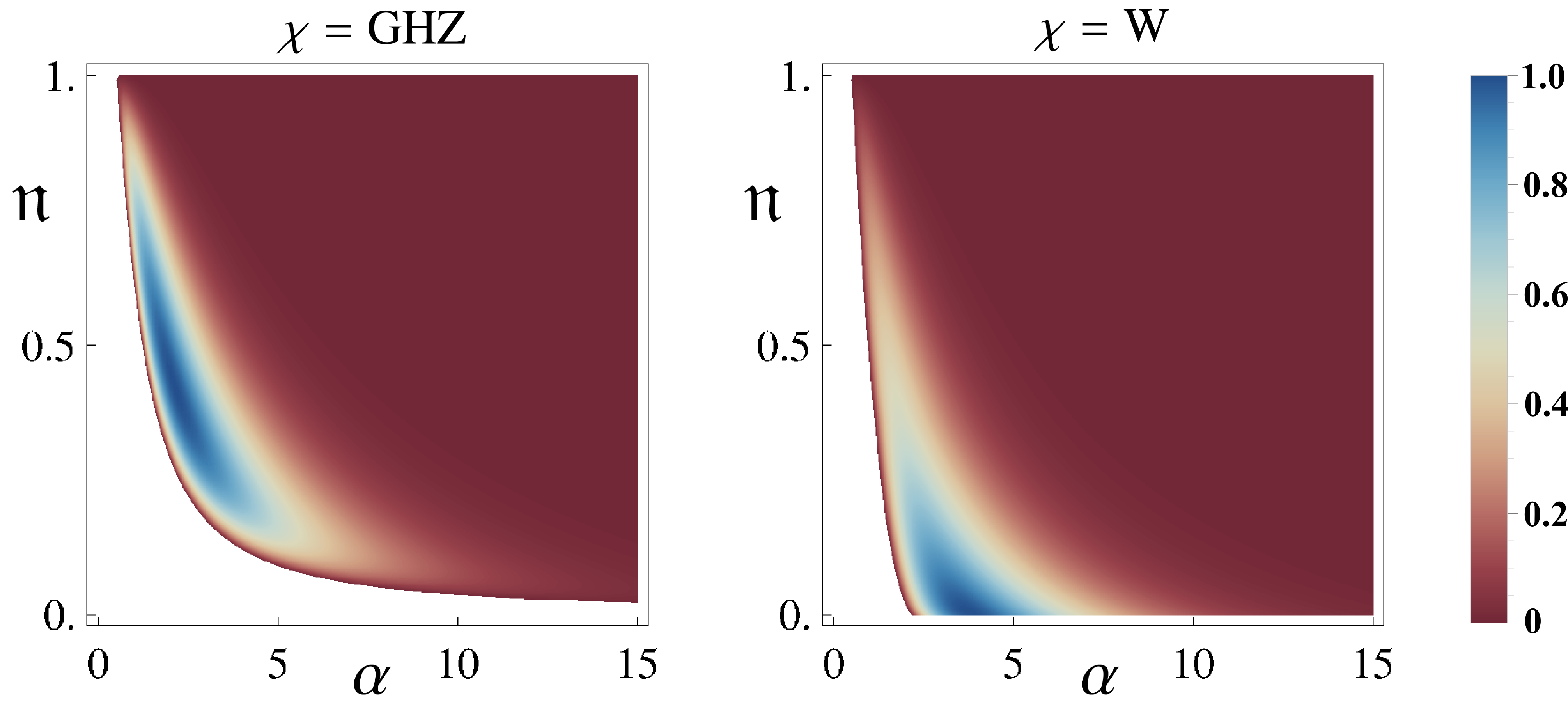}
\caption{\footnotesize Contour plots for the normalized monogamy witness $\delta\mathcal{N}_3^{\alpha}(\rho_\mathfrak{n}^{\chi})/N_\chi$ (see color code), with the respective normalization factors $N_\text{\tiny GHZ}\cong 0.019997$ and $N_\text{\tiny W}\cong 0.073296$, for the state $\rho_\mathfrak{n}^\chi$, as a function of the parameters $\alpha$ and $\mathfrak{n}$. Monogamy holds in the whole colored regions, where $\delta\mathcal{N}_3^\alpha\geq 0$, and does not apply for the pure GHZ state (left panel with $\mathfrak{n}=0$ and $\forall \alpha>0$).}
\label{fig2}
\end{figure}

We start by proving that $\mathcal{N}_3^{\alpha}$ is not monogamous in general. Consider the tripartite GHZ state $\rho_\mathcal{ABC}=\rho_{\mathfrak{n}=0}^\text{\tiny GHZ}$, whose reduced states read $\rho_\mathcal{AB}=\rho_\mathcal{AC}=\left(\ket{00}\bra{00}+\ket{11}\bra{11}\right)/2\equiv\rho_\text{cc}$. We then find $\mathcal{N}_3(\rho_\mathcal{ABC})=\mathcal{N}_2(\rho_\mathcal{AB})=\mathcal{N}_2(\rho_\mathcal{AC})=\ln{2}$, which, for all $\alpha$, can never satisfy the monogamy relation \eqref{monogamy}. Interestingly, however, violations of monogamy for the state $\rho_\mathfrak{n}^\chi$ tend to be relatively rare in the parameter space. To show this, we present in Fig.~\ref{fig2} numerical results for the quantity
\be 
\delta\mathcal{N}_3^{\alpha}(\rho_\mathcal{ABC}):=\mathcal{N}_3^{\alpha}(\rho_\mathcal{ABC})-\left[\mathcal{N}_2^{\alpha}(\rho_\mathcal{AB})+\mathcal{N}_2^{\alpha}(\rho_\mathcal{AC})\right],
\ee 
with $\rho_\mathcal{ABC}=\rho_\mathfrak{n}^\chi$. This is a quantifier that, whenever nonnegative, witnesses monogamy for the measure $\mathcal{N}_3^{\alpha}$ (see the colored region in Fig.~\ref{fig2}). The results show that monogamy is prevented for small values of $\{\mathfrak{n},\alpha\}$ and is saturated for large values. Also, they confirm that the pure GHZ state does not allow for monogamy and that for the pure W state monogamy holds for $\alpha\gtrsim 2.1641$, with $\delta\mathcal{N}_3^\alpha(\rho_{\mathfrak{n}=0}^\text{\tiny W})$ peaking at $\alpha \cong 3.8372$.

\section{Concluding remarks} 
With basis on the criterion of reality proposed in Ref.~\cite{BA2015}, we introduce a bipartite realism-based nonlocality measure that applies for three parts and, from that, a genuine tripartite realism-based nonlocality measure. This quantifier is shown to reduce, for a given class of pure states, to genuine tripartite entanglement and to manifest itself even in cases where no quantum correlations are present. In addition, genuine tripartite realism-based nonlocality turns out to be greatly resistant to noise and deformable into further monogamous quantities for some states. This work paves the way to incursions on the unexplored realm of $n$-partite realism-based nonlocality.

\section*{Acknowledgments} This work was financed in part by the Coordena\c{c}\~ao de Aperfei\c{c}oamento de Pessoal de N\'ivel Superior - Brasil (CAPES) - Finance Code 001. R.M.A acknowledges CNPq (Brazil) and the National Institute for Science and Technology of Quantum Information (INCT-IQ/CNPq).

\end{document}